# Directing the Electrode-Electrolyte Interface Towards Active Nickel-Based Electrocatalysts for Oxygen Evolution Reaction

Ben Wang[1], Tomohiro Fukushima[1], Hiro Minamimoto[2], Andrey Lyalin[1,3,†], Kei Murakoshi[1,‡], Tetsuya Taketsugu[1,4,#]

[1]Department of Chemistry, Faculty of Science, Hokkaido University, Sapporo 060-0810, Japan

[2]Department of Chemical Science and Engineering, Graduate School of Engineering, Kobe University, Kobe 657-8501, Japan

[3]Research Center for Energy and Environmental Materials (GREEN), National Institute for Materials Science, Namiki 1-1, Tsukuba 305-0044, Japan

[4]Institute for Chemical Reaction Design and Discovery (WPI-ICReDD), Hokkaido University, Sapporo 001-0021, Japan

[†]Email: lyalin@icredd.hokudai.ac.jp
[‡]Email: kei@sci.hokudai.ac.jp
[#]Email: t-taketsugu@eis.hokudai.ac.jp

# Abstract

A comprehensive understanding of the electrode-electrolyte interface in energy conversion systems remains challenging due to the complex and multifaceted nature of interfacial processes. This complexity hinders the development of more efficient electrocatalysts. In this work, we propose a hybrid approach to the theoretical description of the OER process on nickel-iron-based oxyhydroxides ($\gamma$-Ni$_{1-x}$Fe$_x$OOH) electrodes in alkaline media as a model system. Multiple reaction pathways represented by the single- and dual-site mechanisms were investigated by taking into account the realistic structure of the catalyst, the doping, and the solvation effects using a simple and computationally feasible strategy. Accounting for the variable solvation effects considerably affects the predicted overpotential in a roughly linear relationship between overpotential and dielectric constant. By incorporating quantum chemical simulations with kinetic modeling, we demonstrate that tuning the local solvation environment can significantly enhance the OER activity, opening new routine ways for elucidation of the emerging issues of OER processes on transition metal oxide surfaces and design of cost-effective, efficient electrocatalytic systems.

# Introduction

Directing energy conversion reactions is a critical challenge for renewable energy technology. Electrochemical water splitting systems involve both the hydrogen evolution reaction (HER) and the oxygen evolution reaction (OER).[1–5] The presence of multiple reaction intermediates in HER and OER processes adds complexity to these reaction networks. These reactions strongly depend not only on the type of electrocatalyst but also on the electrolyte environment, giving the extra flexibility to improve the efficiency of electrochemical energy conversion processes. Notably, the OER is a kinetically slower process in comparison with HER and therefore determines the overall performance of water splitting.[6–8] As such, developing robust design principles for more effective materials is crucial for future advancements.

Considerable efforts have been made for the development of highly active cost-effective electrocatalysts for the OER.[9–11] The most active OER electrocatalysts, however, often rely on expensive precious platinum-group metals (PGMs), such as Ru or Ir, limiting their use in practical applications.[12–15] Therefore, the development of novel catalysts has largely focused on eliminating the dependency on precious metals.[16–24] Currently, cheap and abundant nickel oxyhydroxide NiOOH-based materials doped with transitional metals, such as Fe, are often considered one of the most promising catalysts for water oxidation.[25–37] Despite the large number of works related to the OER activity of the Fe-doped NiOOH-based materials, a complete understanding of the mechanism of the reaction is still elusive with many contradictive interpretations. Thus, even the elucidation of the active sites is problematic, when some studies demonstrate that Fe impurities play a role as the active sites,[25–30] whereas other investigations argue that Ni atoms are the active sites for the OER process on $Ni_{1-x}Fe_xOOH$.[31,32] Therefore, it is crucial to give a deeper understanding of the role of Fe in the enhancement of the OER activity of such materials. However, such investigations should account not only for the properties of electrode materials themselves but also for the properties of electrolytes at the electrified solid-liquid interface.

It is well known that the reaction rate can be tuned by the change of the electrolyte and its composition. Indeed, the electrochemical reactions occur at the interface within the so-called electric double layer with the variable dielectric environments.[38, 39] For example, Bockris *et al*. suggested that in the case of water, $\varepsilon$ decreases from its bulk value of 78.4 to 6 for the inner Helmholtz plane.[39] Recent advanced theoretical studies confirmed a considerable decrease in the values of $\varepsilon$ in the vicinity of solid electrodes.[40] Moreover, the presence of hydrated cations in the solution can affect the local solvation environment and hence influence the overall reaction rate.[40–43] Thus, it was observed experimentally that the cation effect can be explicitly used for modulation of the local solvation environment and the overpotential of the OER, $\eta$, can be modified by changing the cation species of the electrolyte solution.[40] For example, it was demonstrated that the presence of alkali cations ($Li^+$, $Na^+$, $K^+$, and $Cs^+$) in alkaline electrolytes can considerably affect the activity of nickel oxyhydroxide (NiOOH) for the OER.[44] A similar effect has also been observed for OER on $RuO_2$.[45] The overall mechanism of this influence is complex and can depend on several factors, including the interaction of cations with OER intermediates, modulation of the double-layer structure, and structural changes in the catalyst itself due to enhanced intercalation of electrolyte species.[44,46]

In the present work, we propose a simple model accounting for the change in a local solvation environment in the vicinity of a solid-liquid interface by variation of the dielectric constant $\varepsilon$. First, we compare single- and dual-site mechanisms of the OER for the pristine $\gamma$-NiOOH and then examine the Fe-doping effect on the OER activity for the variety of the Fe-impurity position in the parent $\gamma$-NiOOH structure as well as Ni/Fe composition. Using a simple solvation model with the variable dielectric constant we introduce the additional degrees of freedom to optimize the catalyst performance and to model the realistic solvation environment at the solid-liquid interface of the catalyst. Finally, using the obtained theoretical data on the change in the free energy along the reaction pathways calculated for the variety of dielectric constant of the implicit solvent we performed microkinetic simulations of the process to examine the OER activity in a kinetic way. The results of the simulation were discussed by considering the possibility of affecting the processes of multiple proton-electron transfer reactions. Finally, we investigated that even without the catalyst composition optimization, the OER activity can be enhanced by devoting explicit solvation control. Such a simple model allows us to introduce additional degrees of freedom in a simple and computationally affordable way to optimize the complicated multi-electron multi-proton transfer OER process in combination with the traditional descriptors related to the properties of the electrocatalyst itself.

## Results and discussion

### Structural model

To model the active surface area of the $\gamma$-NiOOH ($\gamma$-$Ni_{1-x}Fe_xOOH$) catalyst we adopt the reliable structural model used previously to study OER activity of $\gamma$-$Ni_{1-x}Fe_xOOH$ materials.[35] The $H_2O$ molecules and $K^+$ cations have been intercalated between the $NiO_2$ layers.[47] The bulk γ-NiOOH structure was reoptimized with the set of computational parameters used in the present work. The distance between 2D $NiO_2$ layers in the optimized structure of ~7 Å is in line with the experimental observations.[48–50] The optimized bulk structure has been used for the construction of the (100) surface face as shown in **Figures 1a** and **1b**. Two water molecules have been added to the top layer of the slab to consider the explicit role of the water in the OER mechanism. One of these molecules binds to the $Ni^{4+}$ site, whereas another $H_2O$ molecule spontaneously dissociates on the $Ni^{4+}$ site, with the OH group attached to the $Ni^{4+}$ center and H transferred to the bridging O, as shown in **Figure 1b**. The central unsaturated $Ni^{3+}$ site is considered an active site for the adsorption of the OER intermediates.

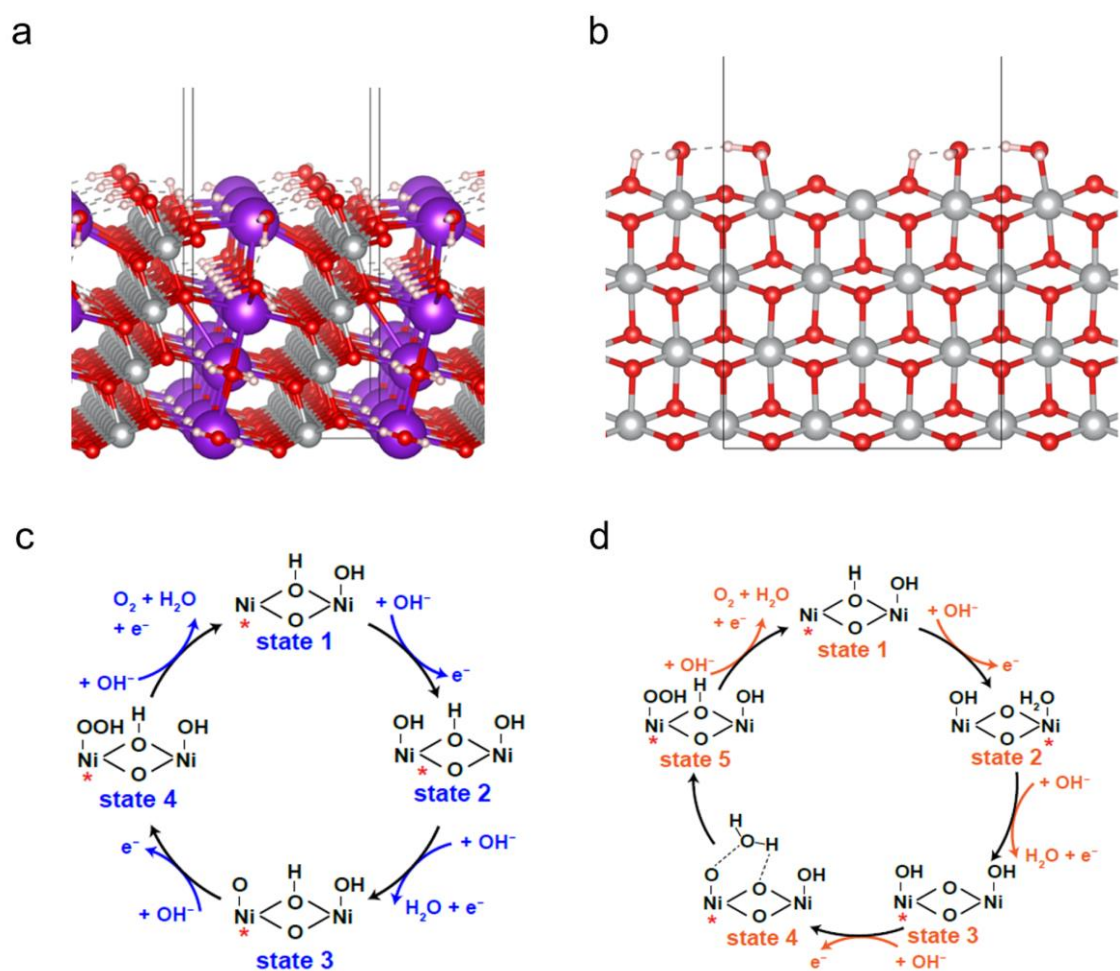

**Figure 1. Structural model and OER mechanisms for the γ-NiOOH catalyst.** The optimized structure of **a** the γ-NiOOH (100) unit cell and **b** the simplified view of the active surface with two explicit water molecules adsorbed on the top layer. Schematic representation of **c** the conventional single-site and **d** the dual-site OER mechanisms on $\gamma$-NiOOH. Ni atoms are colored grey, O atoms are red, H atoms are light pink, and K atoms are purple. The asterisk (*) denotes the active adsorption site.

**OER mechanism in nickel oxyhydroxide: single vs dual active sites**

To evaluate the catalytic activity of the complicated transition metal doped NiOOH structures, one should reveal the possible reaction mechanisms. Due to the presence of multiple reaction intermediates, the OER process can be diversified into multiple reaction paths. The standard OER mechanism consists of four elementary steps, involving the proton and electron. The OER mechanism on metal oxide surfaces can be generally simplified by considering the *OH, *O, *OOH intermediates adsorbed independently on the equivalent active sites, which leads to a universal scaling relation between the binding energies of the OER intermediates.[51]

The OER under alkaline conditions can be written in the following general form:

$$4\text{OH}^- \rightarrow \text{O}_2(g) + 2\text{H}_2\text{O}(l) + 4e^- , \tag{1}$$

where ($g$) and ($l$) refer to the gas and liquid phases, respectively.

The conventional single-site reaction mechanism (**Figure 1c**) consists of four elementary steps:

$$* + \text{OH}^- \rightarrow *\text{OH} + e^- , \tag{2a}$$

$$*\text{OH} + \text{OH}^- \rightarrow *\text{O} + \text{H}_2\text{O}(l) + e^- , \tag{2b}$$

$$*\text{O} + \text{OH}^- \rightarrow *\text{OOH} + e^- , \tag{2c}$$

$$*\text{OOH} + \text{OH}^- \rightarrow * + \text{O}_2(g) + \text{H}_2\text{O}(l) + e^- . \tag{2d}$$

Here the asterisk (*) represents the single active site (the surface Ni atom with the initial formal oxidation state +3) and ∗OH, ∗O, and ∗OOH refer to the species adsorbed on this active site. The change in free energy for each reaction step $n$ of the single-site mechanism is denoted as $\Delta G_{n,\text{single}}$.

In the case of the dual-site OER mechanism (**Figure 1d**), the whole process consists of the following five steps:

$$* + \star\text{OH}\cdots\text{H}^\text{O} + \text{OH}^- \rightarrow *\text{OH} + \star\text{H}_2\text{O} + e^- , \tag{3a}$$

$$*\text{OH} + \star\text{H}_2\text{O} + \text{OH}^- \rightarrow *\text{OH} + \star\text{OH} + \text{H}_2\text{O}(l) + e^- , \tag{3b}$$

$$*\text{OH} + \star\text{OH} + \text{OH}^- \rightarrow *\text{O}\cdots\text{H}_2\text{O}^{(*,\text{O})} + \star\text{OH} + e^- , \tag{3c}$$

$$*\text{O}\cdots\text{H}_2\text{O}^{(*,\text{O})} + \star\text{OH} \rightarrow *\text{OOH} + \star\text{OH}\cdots\text{H}^\text{O} , \tag{3d}$$

$$*\text{OOH} + \star\text{OH}\cdots\text{H}^\text{O} + \text{OH}^- \rightarrow * + \star\text{OH}\cdots\text{H}^\text{O} + \text{O}_2(g) + \text{H}_2\text{O}(l) + e^- . \tag{3e}$$

Here the asterisks (*) and star ($\star$) represent two non-equivalent active sites on the catalyst surface. The active site marked by the star is responsible for the dissociation of the water molecule adsorbed on the surface in the initial configuration and shown as $\star\text{OH}\cdots\text{H}^\text{O}$ to stress that the OH group is attached to the $Ni^{4+}$ site, and H is transferred to the bridging surface O (marked as $H^O$). Note, that in fact the ☆OH···$H^O$ site is also present in the single-site reaction mechanism, however, it does not take part in the overall reaction process, playing the role of a passive spectator. The change in free energy for each reaction step *n* of the dual-site mechanism is denoted as $\Delta G_{n,\,dual}$.

The first step of the dual-site mechanism (3a) consists of the adsorption of the first hydroxyl anion $OH^-$ on the $Ni^{3+}$ active site, with the simultaneous back proton transfer from the bridging O site to the OH group on the $Ni^{4+}$ site, resulting in association of the water molecule ☆$H_2O$. In the second step (3b) the second $OH^-$ anion attacks the ☆$H_2O$ site with deprotonation of the adsorbed water leading to the formation of the ☆OH species and release of the water molecule. Thus, the second step results in the formation of two hydroxyl intermediates *OH and ☆OH adsorbed on the active sites. In the third step (3c) the third $OH^-$ anion attacks the *OH site leading to its deprotonation and formation of the water molecule. However, this water molecule denoted as $H_2O^{(*,O)}$ remains in the vicinity of the formed *O intermediate, bridging the *O site with the surface O atom (**Figure 1d**). Such unique configuration results in the formation of the O-O bond and chemical transformation (3d) of this configuration to a more energetically favorable configuration of *OOH intermediate with a simultaneous proton transfer to the bridging surface O atom, restoring the initial configuration of the second active site ☆OH···$H^O$. Note that this step is purely chemical as it does not include electron release. Finally, in the fifth step (3e) the fourth hydroxyl anion $OH^-$ attacks the *OOH site leading to its deprotonation, the release of the second water molecule, and spontaneous desorption of the $O_2$ molecule.

It is important to note, that the interaction of OER intermediates with $Ni^{3+}$ and $Ni^{4+}$ sites can differ and be modulated independently, potentially breaking the standard scaling relations between the adsorption energies of OER intermediates and lowering the overpotential in a dual-site reaction mechanism.

The formalism presented in Supplementary Information allows us to calculate the changes in free energy $\Delta G_{n,\,single}$ and $\Delta G_{n,\,dual}$ along the considered reaction pathways. **Figures 2a** and **2b** demonstrate the free energy profiles calculated at $U_{RHE} = 0$ V with the use of (S3a)–(S3d) and (S6a)–(S6e) for single- and dual-site OER mechanisms, respectively. Deprotonation of the *OH intermediate is the potential-determining step for both single- and dual-site OER mechanisms, resulting in overpotentials $\eta$ of 1.00 V and 0.68 V, respectively. The 0.3 V decrease in overpotential observed for the dual-site mechanism of OER on *γ*-NiOOH reflects the important role of the binuclear effects accounting for the complex interaction of the intermediates adsorbed on the non-equivalent active sites resulting in the breaking of the standard scaling relation and promotion of the electrocatalytic activity. This non-equivalence of the active sites can be further enhanced by additional doping of the γ-NiOOH electrocatalyst with atoms of transition metals, such as Fe.

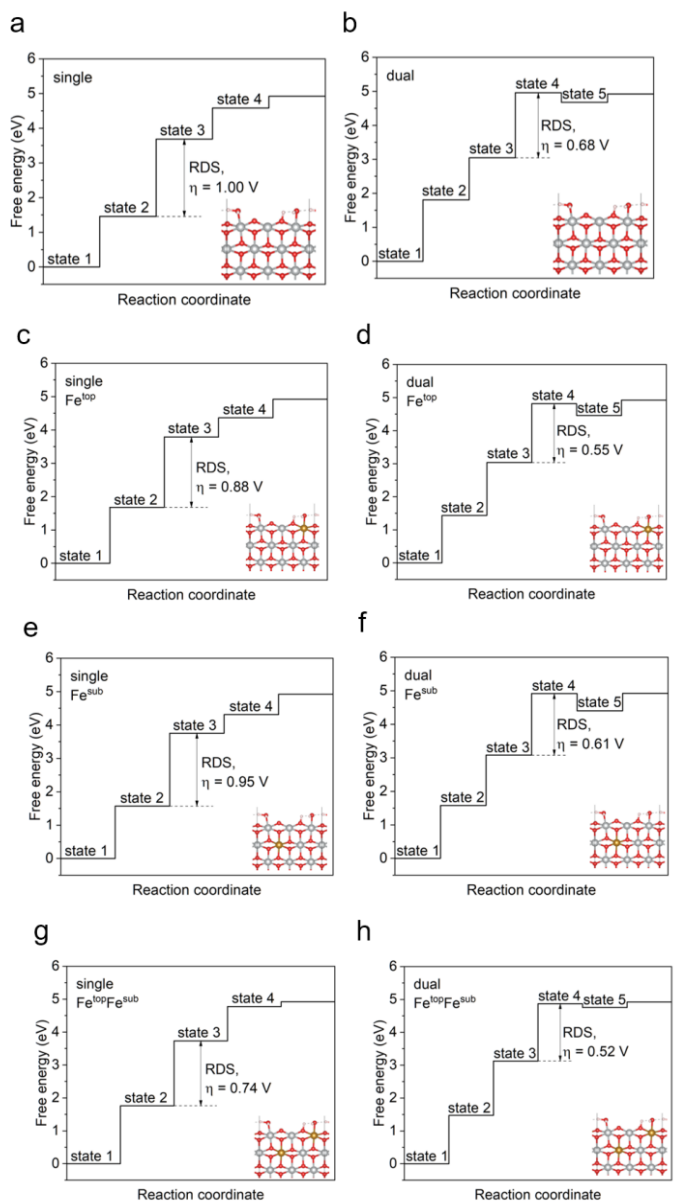

**Figure 2. OER energetics for pure and Fe-doped γ-NiOOH.** Free energy diagrams along the single-site (left) and dual-site (right) OER mechanisms on **a, b** pure γ-NiOOH (100) surface, as well as γ-NiOOH doped with **c, d** one Fe atom in the top layer; **e, f** one Fe atom in the first sublayer, and **g, h** two Fe atoms in the top and sublayer. All pathways are calculated at $U = 0$ $V_{RHE}$. The potential-determining step is indicated by the vertical arrow along with the calculated value of the overpotential $\eta$. Ni atoms are colored grey, Fe atoms are brown, O atoms are red, H atoms are light pink, and K atoms are purple.

**Influence of Fe doping on OER activity**

To elucidate the role of Fe doping on the OER process, we have constructed a number of the structural models substituting $Ni^{4+}$ sites in the parent γ-NiOOH structure by Fe in the following non-equivalent positions: (i) $Ni^{4+}$ site in the top layer with the adsorbed OH group after $H_2O$ dissociation ($Fe^{top}$); (ii) $Ni^{4+}$ site in the sublayer in the close vicinity of the surface active sites ($Fe^{sub}$); (iii) both of the above sites ($Fe^{top}Fe^{sub}$). Indeed, previously it has been suggested that Fe atoms preferably substitute $Ni^{4+}$ sites in γ-NiOOH.[35] The calculated formation energies of the Fe-doped structures (i)−(iii) are −2.07 eV, −1.34 eV, and −1.67 eV per Fe atom, respectively. We have also confirmed that the formation energy for the substitution of $Ni^{3+}$ in the first surface layer is −0.82 eV, which is energetically less favorable than the substitution of the $Ni^{4+}$ sites.

Similarly, in comparison with the pure Fe-free γ-NiOOH structure, the theoretical overpotential of OER was evaluated as shown in **Figures 2c – 2h**. The overpotential is reduced to 0.88 V (single-site), and 0.55 V (dual-site) for the single Fe atom doped in the first layer, showing the better OER activity of the Fe-doped system. The corresponding overpotentials calculated for the single- and dual-site pathways, for a system with a single Fe atom doped in the second layer are 0.95 V and 0.61 V, respectively. The increase in the concentration of Fe dopants up to two atoms per unit cell, results in a further decrease in the overpotential to 0.74 V and 0.52 V calculated for single-site and dual-site pathways, respectively. The decrease in overpotential of OER as a result of Fe doping demonstrates the synergistic effect between Fe and Ni leading to the better catalytic performance of γ-$Ni_{1−x}Fe_xOOH$.

The results of calculations of OER overpotential for γ-$Ni_{1−x}Fe_xOOH$ catalyst as a function of Fe concentration are summarized in **Figure S1a**. The dual-site mechanism of the reaction is considered. As was already mentioned, $η$ decreases up to 0.52 V almost linearly, with an increase in Fe concentration up to two atoms per unit cell. This concentration of Fe atoms corresponds to 17% of Fe in the whole unit cell or 33% of Fe in the surface layer. To gain deeper insights into the mechanism of the enhancement of OER activity due to Fe doping, we have performed a Bader charge analysis,[52,53] which allowed us to explore how Fe doping affects the local electronic environment at the atomic level. Our calculations of the distribution of Bader charges demonstrate that the bridging O atoms connecting the doped Fe atoms additionally gain −0.1|e| negative charge. As a result of Fe doping more electrons have been shared by the bridging O sites, and the excess of the electron density around the active O site that bridges Ni and Fe increases (**Figure S1b**). This in turn can lead to an increase in the electron charge density around the active Ni site. Therefore, the reducibility of this Ni site can be enhanced due to its higher electron affinity, affecting its ability to adsorb OER intermediates. Thus, using Fe doping it is possible to modify the local electronic properties of the active sites, promoting the OER activity of $Ni_{1-x}Fe_xOOH$ electrocatalyst. However, the structural and electronic properties of the catalytic surface are not the only factors governing the OER activity.

**Influence of implicit solvation on OER activity**

The calculations performed in the previous section have been performed without accounting for the solvation effects. As was already mentioned in the introduction, electrocatalytic reactions occur in the electrified solid-liquid interface region.[40,54,55] To describe the solvation effects on OER activity in a simple model way, we have adopted an implicit solvent approach of a polarizable continuum model (PCM) with a variable scalar dielectric constant, $\varepsilon$.[56] Optimization of all adsorbed OER intermediates on the Fe free γ-NiOOH has been performed for a number of dielectric constants ε variable in the range from 1 to 120, both for single-site and dual-site mechanisms. This modulation reflects the fact that the dielectric constant in the solid-liquid interface region can considerably deviate from its bulk value.[40,57−59] The solvation corrections to the free energies, $\Delta G_{sol}(\varepsilon)$, have been obtained as a difference in free energies of intermediates $G(\varepsilon)$-$G(0)$ calculated with and without accounting for the solvent environment. For the sake of simplicity, the same values of $\Delta G_{sol}(\varepsilon)$ have been applied for the Fe-doped γ-$Ni_{1-x}Fe_xOOH$ systems due to the minor influence of the doping on solvation energies of OER intermediates.[38,60−62]

Let us now analyze how the solvation effects accounted for within a simple PCM model influence the OER activity of γ-$Ni_{1-x}Fe_xOOH$ structure with two Fe dopants. **Figure 3** demonstrates the dependence of (a, b) free energies of OER intermediates, and (c, d) overpotentials calculated for the single-site (left column) and dual-site (right column) reaction mechanisms, as a function of the dielectric constant $\varepsilon$. As one can see from **Figure 3** free energies of OER intermediates decrease with an increase ε from 1 to 120, both for single-site and dual-site reaction mechanisms. This effect can be easily understood, as solvation effects stabilize the OER intermediates. The free energy of OER intermediates varies in the range of 0.5 eV in the considered diapason of $\varepsilon$, pointing out the importance of the solvation effects on OER activity. In both cases of the single-site and dual-site mechanisms oxidation of the adsorbed *OH intermediate is the potential-determining step in the whole range of the considered $\varepsilon$, as indicated by the vertical arrows in **Figures 3a** and **3b**.

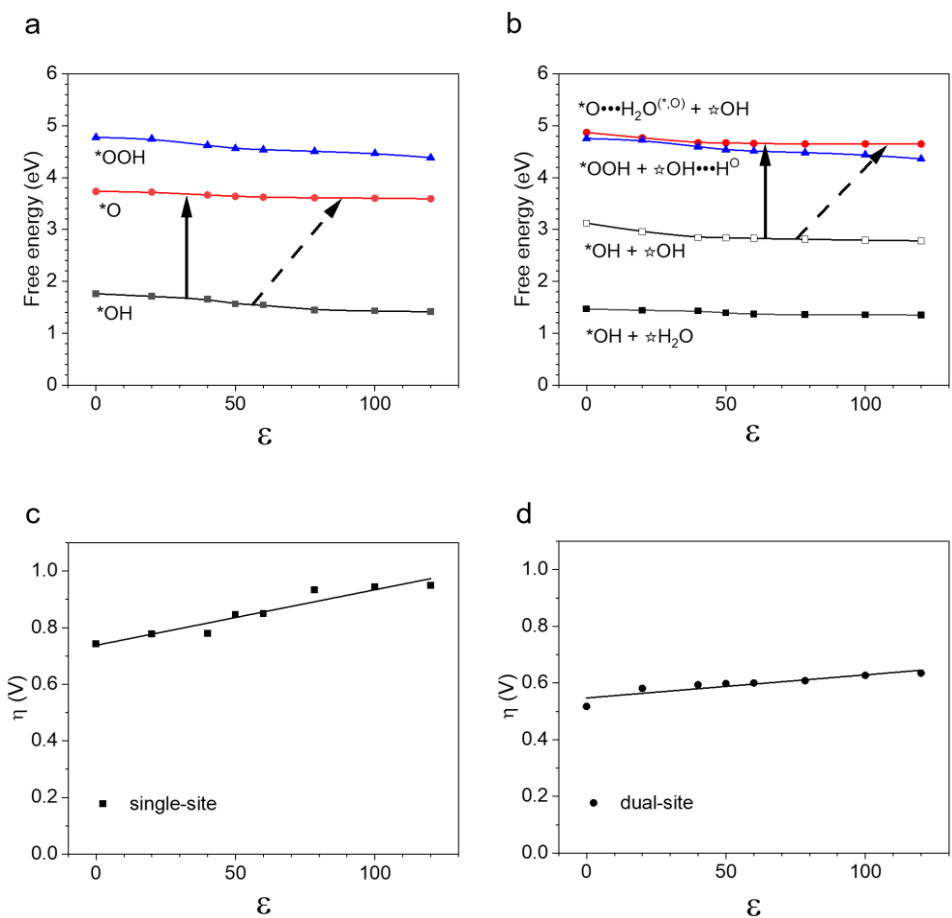

**Figure 3. Role of the dielectric environment. a, b** Free energy for the OER intermediates along the reaction pathway at 0 $V_{RHE}$, and **c, d** values of the overpotential $\eta$ as a function of the dielectric constant $\varepsilon$, calculated for the single-site (left) and dual-site (right) mechanisms, respectively. The potential-determining transitions are marked by the solid line arrow (transition in the fixed dielectric environment) and the dashed line arrow (transition, where the reaction intermediates are in the different dielectric environment). Two Fe atom doped $\gamma$-NiOOH (100) surface is considered.

The dependence of the change of free energy for each reaction step $\Delta G_n$ on $\varepsilon$ is more complicated (**Figure S2**). We have found that steps related to the adsorption of OH⁻ on the active sites show a decrease in $\Delta G_n$, while steps related to the production of $H_2O$ led to an increase in corresponding $\Delta G_n$ with an increase in $\varepsilon$. Such behavior of $\Delta G_n$ considerably affects the values of overpotential as a function of $\varepsilon$. Our calculations demonstrate, that overpotential is very sensitive to the values of $\varepsilon$ and increasing with an increase in dielectric constant. Thus, the values of the overpotential $\eta$ calculated for single- and double-site reaction mechanisms are 0.74 V and 0.52 V for $\varepsilon = 1$, increasing up to 0.93 and 0.61 for $\varepsilon$ = 78.4 corresponding to a dielectric constant of liquid water at ambient condition (**Figures 3c** and **3d**). The important issue is, however, the careful choice of the $\varepsilon$ value. As it was already mentioned the electrochemical processes occur at the solid-liquid interface region, where the dielectric constant $\varepsilon$ considerably deviates from its bulk values. Therefore, simple estimation of solvation effects using implicit models like PCM with the dielectric constant of bulk water can lead to erroneous results and considerable overestimation of the value of overpotential over 0.1 V. Obviously, modeling solvation environment in the double-layer region requires more accurate theoretical models than PCM approach employed here. However, our model with the variable $\varepsilon$ is very useful to understand on the qualitative level the role of solvation effects on the OER process, simultaneously considering a large variety of the vital factors affecting OER, such as the structure and composition of the electrocatalyst itself and the details of the reaction mechanism, accounting for the complicated dual-site pathways.

**Kinetic simulations**

Based on the calculated free energy of OER intermediates for the variety of the dielectric constants of the implicit solvent, we construct a microkinetic model, opening a way for the direct comparison of theoretical predictions with the experimental observables, such as I-V curves and Tafel slopes. Such an approach can explicitly visualize the relative role of theoretical descriptors in electrocatalysis and assist with the elucidation of the rate-determining steps. In this work, we adopt the microkinetic model for the coverage-dependent current density, proposed by Takanabe et al.[63] As follows from our quantum-chemical calculations, in the case of the single-site mechanism the reaction (2b) determines the overall reaction rate. Therefore, the kinetic current, *J*, can be described according to Ref. [**63**] as

$$J = nFA\, k_{2b}^0\, \theta_1\, a_{OH^-}\, e^{(1-\alpha)f\eta_{2b}}, \qquad (4)$$

where *n* is the number of electrons involved, *F* is the Faraday's constant (*F*= 96485 C mol$^{-1}$), *A* is the reaction site density of the catalytic surface, $k_i^0$ is the standard rate constant, $\eta_i$ is the overpotential (the difference between the electrode and the theoretical value of the standard potential, obtained from the calculated free energies of the reaction intermediates) for the reaction *i*, $a_{OH^-}$ is the hydroxide ion activity (effective concentration of the hydroxide ion in the solution), $\alpha$ is the electron transfer coefficient which we set to be equal 0.5 for simplicity, *f* denotes the ratio *F/RT*, where *R* is the universal gas constant equal to 8.314 J mol$^{-1}$ K$^{-1}$, and *T* = 298.15 K is the temperature. The surface coverage $\theta_1$ by *OH intermediates can be expressed as follows:

$$\theta_1 = \frac{\widehat{K}_{2a}^0 a_{OH^-}\, e^{f\eta_{2a}}}{1+\widehat{K}_{2a}^0 a_{OH^-}\, e^{f\eta_{2a}}}, \tag{5}$$

where $\widehat{K}_i^0$ defines the ratio $k_i^0/k_{-i}^0$. The coverage of the empty sites, $\theta_0$, can be expressed as

$$\theta_0 = 1 - \theta_1. \tag{6}$$

For the dual-site mechanism the third $OH^-$ transfer, the reaction (3c), determines the overall reaction rate. However, in this case, the formalism described in Ref. [63] cannot be used directly to calculate the current density, because two catalytic sites are involved in the process. While the first (3a), third (3c), and fourth (3e) $OH^-$ attack occurs at one site (A), and the second step (3b) occurs at the second site (B). Therefore, the current density can be expressed as follows:

$$J = nFA\, k_{3c}^0\, \theta_1^A\, a_{OH^-}\, e^{(1-\alpha)f\eta_{3c}}, \tag{7}$$

where $\theta_1^A$ is the coverage of the site A by *OH intermediates:

$$\theta_1^A = \frac{\widehat{K}_{3a}^0 a_{OH^-}\, e^{f\eta_{3a}}}{1+\widehat{K}_{3a}^0 a_{OH^-}\, e^{f\eta_{3a}}}, \tag{8}$$

and coverage of the empty sites A is defined as

$$\theta_0^A = 1 - \theta_1^A. \tag{9}$$

The expressions (7)-(9) are obtained using the assumption that sites A and B are independent, and coverage of site B does not affect site A. The derivation of Eqs. (4)-(9) is given in Supplementary Information. Note, that the kinetic currents (4) and (7) depend on the potential-dependent coverage of the intermediate species, related to their formation and consumption rates. The above equations depend on the kinetic rate constants for the reactions (2a)-(2b) and (3a)-(3c), which are not known in advance and can be used as arbitrary parameters or obtained by the fitting the theoretical current density to the experimental I-V curve. In the present work in the case of the single-site reaction mechanism, the current density (4) is calculated using the following values $k_{2a}^0/k_{2b}^0/k_{-2a}^0 = 10^2/10^7/1$ and $a_{OH^-} = 1$ as suggested in Ref. [63]. In the case of the dual-site reaction mechanism, the expression (7) should more adequately describe the experimental current density. Therefore, in this case, we obtain the kinetic constants by fitting theoretical current density (7) to the typical experimental values for the pure NiOOH catalyst reported in the classical work by Corrigan[64]. Such fitting results in the following values $k_{3a}^0/k_{3c}^0/k_{-3a}^0 = 7.5 \times 10^3/2.6 \times 10^4/1$.

      **Figures 4a** and **4b** demonstrate the results of I-V curves simulations for the single- and dual-site reaction mechanisms in vacuum, respectively. Note that for simplicity we use the same set of kinetic constants for the pure and Fe-doped samples. As one can see from **Figure 4a** use of the kinetic constants suggested in Ref. [63], allows us

to describe the potential dependence of the current density for the large currents (≥20 mA/cm$^2$), demonstrating the increase in the OER activity of the considered structures as pure γ-NiOOH < Fe$^{sub}$ < Fe$^{top}$ < Fe$^{top}$ Fe$^{sub}$, similar to the model entirely based on the analysis of the change in free energies (see Figure 2). A qualitatively similar increase in OER activity of NiOOH catalysts with an increase in the concentration of the Fe dopant was observed experimentally.[64,65] However, at the current densities below 10 mA/cm$^2$, **Figure 4a** demonstrates noticeable pre-equilibrium effects due to the coverage dependence of the Tafel-slopes.[63] Interestingly, attempts to fit the theoretical I-V curve for the pure NiOOH catalyst to the experimental values result in very strong pre-equilibrium effects and considerable deviation from experimental observations for Fe-doped samples.[64,65] Therefore, the single-site model cannot describe adequately the OER kinetics for the Fe-doped NiOOH-based catalysts with the unified set of the kinetics constants.

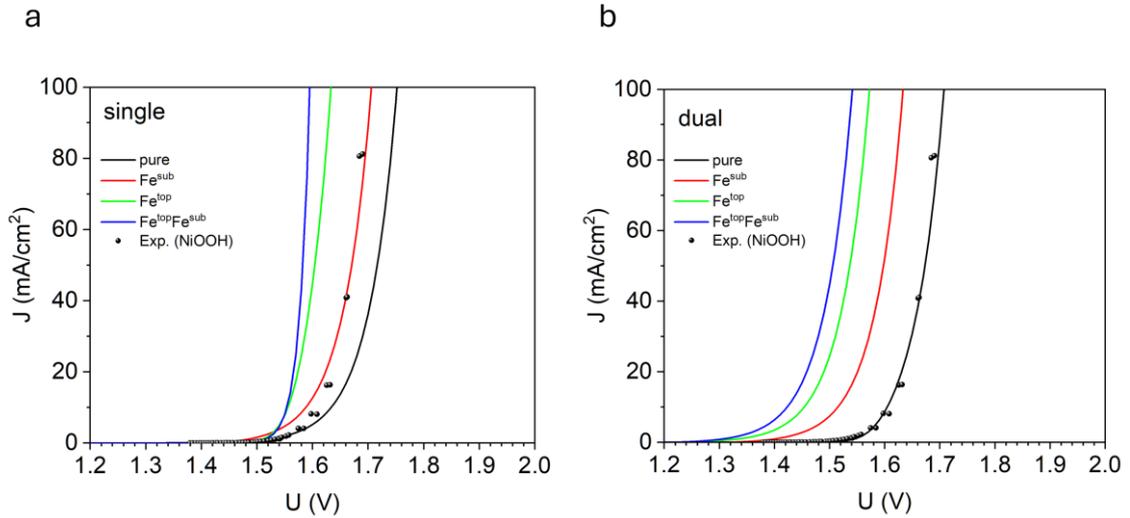

**Figure 4**. **Kinetic simulations for pure and Fe-doped γ-NiOOH**. I-V curves modeled for (a) the single-site, Eq. (4), and (b) dual-site, Eq. (7), OER mechanisms on the pure and Fe-doped γ-NiOOH. Black curve – pure γ-NiOOH, red curve – one Fe atom doped in the first sublayer, green curve – one Fe atom doped in the top layer, blue curve – two Fe atoms in the top and sublayer, black dots – experimental data the pure NiOOH taken from Ref. [64]. Calculations are performed for $\varepsilon = 1$.

On the other hand, in the case of the dual-site mechanism, we have achieved excellent fitting of the theoretical I-V curve for the pure γ-NiOOH sample with the experiment (**Figure 4b**) as well as the qualitatively correct behavior of OER activity with Fe doping (increase in activity with Fe doping). For example, in the case of the dual-site mechanism the kinetic overpotential at 10 mA/cm² decreased from 0.38 V (γ-NiOOH) to 0.29 V (γ-NiOOH with Fe dopants in sub layer), to 0.23 V (γ-NiOOH with Fe dopants in top layer), and to 0.20 V (γ-NiOOH with Fe dopants both in top and sub layers). These values are in good accord with the overpotentials reported for various $Ni_{1-x}Fe_xOOH$ catalysts, which are typically scattered in the range of 0.26−0.48 V.[35] It is also important to note that not only Fe concentration but also the position of the doped atom impacts the catalytic activity. For example, the OER activity is larger for the Fe impurity in the top layer if compared with the sub-layer impurity, because the surface site plays an important role in the activity.

Overall, our kinetic analysis demonstrates that the dual-site reaction mechanism more adequately describes the OER process on the complicated NiOOH-based catalysts, possessing non-equivalent catalytic sites. Therefore, further analysis of the solvation environment on the OER kinetics we perform only for the dual-site reaction mechanism.

**Effect of the solvation environment on OER kinetics**

Theoretical results presented in **Figure 4** were obtained without accounting for the solvation environment, when $\varepsilon = 1$. Below we show how the solvation environment affects the OER kinetics. In the case of the dual-site mechanism, free energies of *OH + ☆OH and *O···H$_2$O$^{(*,O)}$ + ☆OH intermediates define the RDS (**Figure 3b**). Here we model the kinetic current with the independent variation of the dielectric constants $\varepsilon'$ and $\varepsilon''$ for the initial *OH + ☆OH and *O···H$_2$O$^{(*,O)}$ + ☆OH intermediates, respectively. We also analyze the standard approach, when all OER intermediates are embedded in the same dielectric environment, i.e., $\varepsilon' = \varepsilon'' = \varepsilon$ (vertical transition in **Figure 3b**).

**Figures 5a – 5d** demonstrate the potential dependent current-density calculated for dual-site OER mechanism for **(a)** the pure γ-NiOOH catalyst, and γ-NiOOH catalyst with **(b)** Fe$^{sub}$, **(c)** Fe$^{top}$, and **(d)** Fe$^{top}$Fe$^{sub}$ impurities. Dashed curves present the case when $\varepsilon' = \varepsilon'' = \varepsilon = 1$, blue dots correspond to the case of the bulk water $\varepsilon' = \varepsilon'' = \varepsilon = 78.4$, and dashed-dotted curves $\varepsilon' = \varepsilon'' = \varepsilon = 120$. The possible spread of the I-V curves as a result of variation of the dielectric constant ε from 1 to 120 is shown by the dark grey area. As one can see from Figure 5, the dielectric environment considerably affects the OER kinetics. The decrease in the dielectric constant results in an increase in the OER activity and a decrease in the kinetic overpotential. Thus, for example for the pure γ-NiOOH catalyst the kinetic overpotential at 10 mA/cm² decreases from 0.45 V to 0.38 V. A similar effect is observed for the Fe-doped γ-NiOOH systems. Therefore, modulation of the dielectric environment can be used to boost the OER activity of the γ-NiOOH based catalysts. Moreover, independent variation of the dielectric constants $1 < \varepsilon'$ 120 and $1 < \varepsilon'' < 120$, when the reaction intermediates are in the different dielectric environment (see

**Figure 3b**), results in a very large spread of I-V curves in the area shown by light-grey color in **Figures 5a – 5d**. Thus, even for the pure γ-NiOOH catalyst one can achieve the kinetic overpotential at 10 mA/cm² to be 0.2 V.

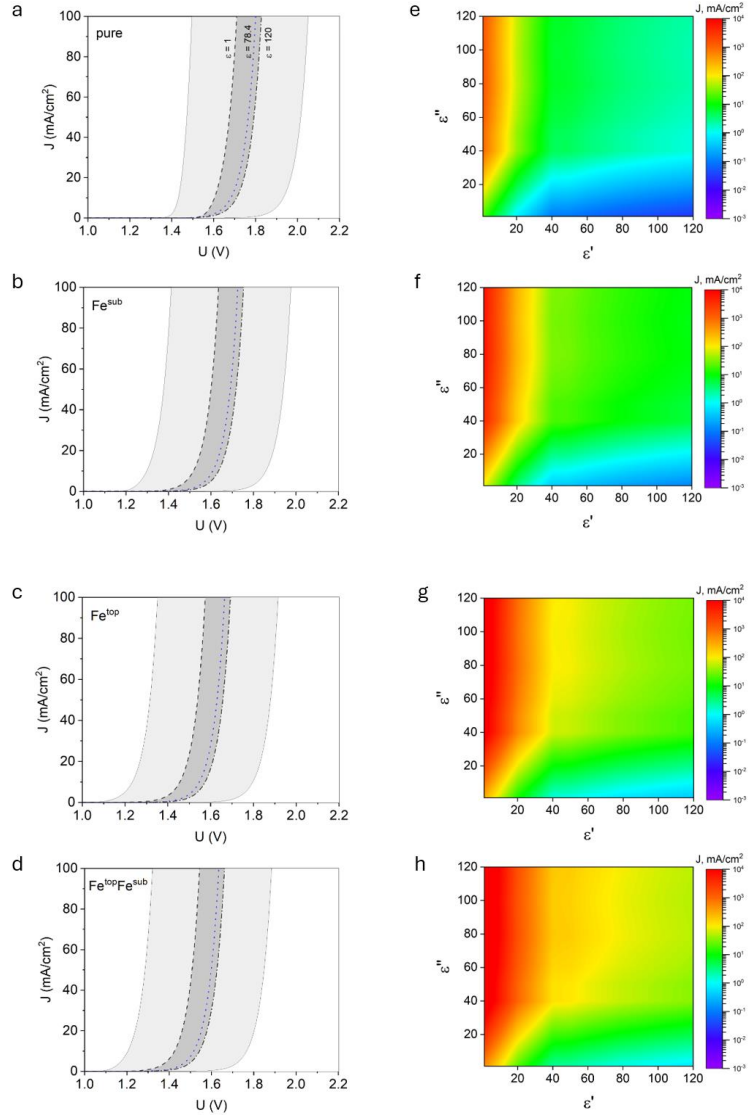

**Figure 5. Effect of the dielectric environment on OER kinetics. (a)-(d)** Simulated I-V curves for the dual-site OER mechanisms on the pure and Fe-doped γ-NiOOH for the variety of dielectric constants. Dashed curves – $\varepsilon' = \varepsilon'' = \varepsilon = 1$, blue dots – $\varepsilon' = \varepsilon'' = \varepsilon = 78.4$, dashed dotted curves – $\varepsilon' = \varepsilon'' = \varepsilon = 120$. The dark grey area corresponds to the possible spread of I-V curves for variation of dielectric constant $\varepsilon$ from 1 to 120. The light grey area corresponds to the possible spread of I-V curves in the case of the independent variation of $\varepsilon'$ and $\varepsilon''$. **(e)-(f)** 2D maps of the current density as a function of $\varepsilon'$ and $\varepsilon''$ for the pure and Fe-doped γ-NiOOH calculated at $U = 1.63$ V for the dual-site OER mechanism.

**Figures 5e – 5h** demonstrate the 2D map of the current density as a function of $\varepsilon'$ and $\varepsilon''$ for the pure and Fe-doped γ-NiOOH calculated at $U = 1.63$ V for the dual-site OER mechanism. The calculated current density is very sensitive to the solvation environment of the *OH + ☆OH intermediate, $\varepsilon'$, considerably increasing for the relatively small values of $\varepsilon' < 20$. On the other hand, the current density is less sensitive to the dielectric environment of the *O⋯H$_2$O$^{(*,O)}$ + ☆OH intermediate, $\varepsilon''$, allowing the wide range of $\varepsilon'' > 20$ for the pure γ-NiOOH catalyst (**Figure 5e**) and even wider possible range of $\varepsilon''$ for the high kinetic current in the case of Fe-doped γ-NiOOH catalysts (**Figures 5f – 5h**). This is a very interesting result, showing that the high OER activity can be potentially achieved not only by the modification of the catalyst itself but by the independent modulation of the solvation environment for the different reaction intermediates.

Usually, the activity of the catalyst can be controlled by the catalyst composition and doping affecting the binding energy of the reaction intermediates. If the catalytically active sites remain unchanged after the doping by additional elements, the alteration of the binding energy of the reaction intermediates can be explained by the charge re-distribution in the catalyst surface due to the presence of the orbital interaction, electronic conjugation, etc. However, the implicit solvation effects at the electrode-electrolyte interface can also affect the stability of the reaction intermediates. In this study, the observed difference in reaction kinetics is explained by the modification of the binding energy of the intermediates for the rate-determining step due to several factors. It is important, that such modification can be performed only by variation of the dielectric environment for the reaction intermediates, leading to considerable enhancement of the OER activity. Therefore, the modification of the dielectric behavior during the reaction steps in the multiple proton-coupled electron transfer reactions can be validated for the further development of the catalysts and/or the dielectric environments. Recent vibrational spectroscopic studies on water molecules prove that autoprotolysis processes of water can vary depending on the freedom of proton mobility at electrified interfaces.[66,67] One can expect that active control for the increments in proton mobility of water molecules at the specific OER intermediates may increase the apparent dielectric constant of water, leading to improvements in OER activity.

From the aforementioned discussion, we can envision that the screening of the dielectric environments is utilized for the simulation of the activity of the electrochemical reactions. In particular, in the case of a charge-transfer step (either for proton-transfer or electron-transfer), the dynamic charge re-distribution at the solid/liquid interface of the catalyst surface can occur. The dynamic modulation in the dielectric environment dictates the efficient pathway without energy loss. Generally, it has been believed that the surface composition is important for the determination of the free energy of the reaction intermediates and the resulting OER behavior. However, real catalyst surfaces have various surface groups during the OER catalysis, such as *OH, *O, *OOH, and a combination of them. Due to the variety in the surface structure, the solvation environment of the electrode surface in the double-layer structure can be variable for the reaction intermediates.[68] In addition, we experimentally observed the anomalously high OER activity from the nanostructured electrode.[69,70] Therefore, our demonstration of the independent combination of the dielectric constant modulation can be utilized for the real-surface description of the catalysts.

## Summary

This study has focused on the extent to which comprehensive quantum chemical calculations combined with kinetic modeling-based simulations can be used for declaring experimental operation factors in a theoretical way.

We propose a hybrid approach to the theoretical description of the OER process simultaneously considering a variety of key factors influencing the OER process, such as the structure and composition of the catalyst, the complexity of the reaction mechanisms accounting for the complicated dual-site reaction pathways, and the change in a local solvation environment with a variable dielectric constant in a simple and computationally feasible manner. We have demonstrated that the dual-site reaction mechanism taking advantage of the synergy between Fe and Ni is more favorable than the single-site one, which is consistent with previous observations. The solvation effects of the implicit solvent considerably affect the predicted overpotential, and a roughly linear relationship between overpotential and dielectric constant has been found. Our simple estimations demonstrate that the use of implicit solvation models with the dielectric constant of liquid water ($\varepsilon$ = 78.4) can lead to erroneous results and considerable overestimation of the OER overpotential over 0.1 V. Using the robust kinetic modeling simulation tool, we have further examined the solvation effect on OER activity in a microkinetic manner. It is demonstrated that a desired increase in the OER activity, which is usually achieved by changing the composition of the catalyst can be attained by the modulation of the dielectric environment. These findings demonstrate that the complex combined quantum chemical and kinetic modeling simulations can open a new routine approach for elucidation the emerging issues of OER processes on transition metal oxide surfaces in a simple and computationally affordable way and achieve a deeper understanding of OER mechanisms.

# Methods

The spin-polarized density-functional theory (DFT) calculations have been performed using the Vienna ab initio Simulation Package (VASP).[71,72] The projector-augmented wave (PAW) method[73] and Perdue-Burke-Ernzerhof (PBE) exchange-correlation functional[74] with the Hubbard U corrections within the rotationally invariant Dudarev approach[75] have been used. The U parameter for 3d orbitals of Ni and Fe has been selected in accord with the previous calculations to be equal to 5.5 eV and 3.3 eV, respectively.[76–78] The Kohn-Sham wave functions have been expanded in a plane wave basis set with a kinetic cutoff energy of 540 eV. Dispersion terms have been introduced using the D3 Grimme's parametrization.[79] The vibrational frequency calculations have been performed to obtain the zero-point energy (ZPE) corrections. The entropic contributions to the free energies have been obtained using the ideal gas approximation[80] for the free molecules and considered to be negligible for the adsorbed species assuming the main contribution to the entropy from the translational degrees of freedom in accord with our previous works.[81,82]

To model OER in the liquid phase, we considered the effect of including the implicit water solvation via a dielectric polarizable continuum model (PCM) by solving the linearized Poisson−Boltzmann equation as implemented in VASPsol.[83,84] In this self-consistent continuum dielectric model, the relative permittivity $\varepsilon$ of solvent has been set to 20, 40, 50, 60, 78.4 (water), 100, and 120. The following parameters have been used for the width of the dielectric cavity $\sigma = 0.6$, a cutoff charge density $\rho = 0.00025$ Å$^{-3}$, and an effective cavity surface tension value of 0.6 meV/Å$^2$.

The structural model for the $\gamma$-Ni$_{1-x}$Fe$_x$OOH catalyst used in this work has been chosen based on the highly reliable theoretical model proposed by Ceder et al.[47] This model suggests the chemical formula K$_{1/3}$(H$_2$O)$_{2/3}$Ni$_{1-x}$Fe$_x$O$_2$, of the structure consisting of the 2D Ni$_{1-x}$Fe$_x$O$_2$ layers intercalated with K$^+$ cations and H$_2$O molecules. It gives the reliable interplanar spacing of 6.8 Å, as well as the proper averaged mixed formal oxidation state close to +3.67 for transition metal elements in $\gamma$-Ni$_{1-x}$Fe$_x$OOH-like catalyst, consistent with previous studies,[25,29,33,35] and has been used recently for the sophisticated analysis of the OER activity.[35] The OER has been considered to take place at the (100) surface plane of the $\gamma$-Ni$_{1-x}$Fe$_x$OOH structure which possesses high catalytic activity.[85,86] The 3 × 1 surface unit cell slab containing four Ni$_{1-x}$Fe$_x$O$_2$ layers has been used to model the surface. The two bottom layers were fixed at the bulk positions, while the two top layers, including adsorbates, were allowed to fully relax. A vacuum region of 15 Å was created to ensure negligible interaction between the periodically replicated images. The Brillouin zone was sampled by the Monkhorst-Pack 3 × 3 × 1 *k*-point mesh to optimize the structure, and a more dense 5 × 5 × 1 mesh was used to obtain the energy.[87] A Gaussian electronic smearing of width 0.01 eV was used for atomic relaxation and vibrational frequency calculations. An absolute force threshold of 0.01 eV/Å was imposed during structural relaxation. All images are generated by VESTA.[88]

## ASSOCIATED CONTENT
**Supplementary Information**

Additional computational methods; calculation of solvation correction on free energy, corrected binding energy, energy difference, and theoretical potential; more details about the microkinetic modeling simulation.

**Acknowledgments**

This work was partly supported by GteX Program, Japan, Grant Number JPMJGX23H2, and MEXT Program: Data Creation and Utilization-Type Material Research and Development Project Grant Number JPMXP1122712807. Calculations were performed using computational resources of the Institute for Solid State Physics, the University of Tokyo, Japan; the Numerical Materials Simulator, NIMS, Tsukuba, Japan; and the Research Center for Computational Science, Okazaki, Japan (Project: 23-IMS-C016).

**References**

1. Chatenet, M.; Pollet, B. G.; Dekel, D. R.; Dionigi, F.; Deseure, J.; Millet, P.; Braatz, R. D.; Bazant, M. Z.; Eikerling, M.; Staffell, I.; et al. Water Electrolysis: from Textbook Knowledge to the Latest Scientific Strategies and Industrial Developments. *Chem. Soc. Rev.* **2022**, *51*, 4583−4762.
2. Ghosh, S.; Basu, R. N. Multifunctional Nanostructured Electrocatalysts for Energy Conversion and Storage: Current Status and Perspectives. *Nanoscale* **2018**, *10*, 11241−11280.
3. Wang, M.; Zhang, L.; He, Y.; Zhu, H. Recent Advances in Transition-metal-sulfide-based Bifunctional Electrocatalysts for Overall Water Splitting. *J. Mater. Chem. A* **2021**, *9*, 5320−5363.
4. He, R.; Huang, X.; Feng, L. Recent Progress in Transition-Metal Sulfide Catalyst Regulation for Improved Oxygen Evolution Reaction. *Energy Fuels* **2022**, *36*, 6675−6694.
5. Yuan, C.-Z.; Hui, K. S.; Yin, H.; Zhu, S.; Zhang, J.; Wu, X.-L.; Hong, X.; Zhou, W.;

Fan, X.; Bin, F.; Chen, F.; Hui, K. N. Regulating Intrinsic Electronic Structures of Transition-Metal-Based Catalysts and the Potential Applications for Electrocatalytic Water Splitting. *ACS Mater. Lett.* **2021**, *3*, 752−780.

6. Song, F.; Bai, L.; Moysiadou, A.; Lee, S.; Hu, C.; Liardet, L.; Hu, X. Transition Metal Oxides as Electrocatalysts for the Oxygen Evolution Reaction in Alkaline Solutions: An Application-Inspired Renaissance. *J. Am. Chem. Soc.* **2018**, *140*, 7748−7759.

7. Suen, N. T.; Hung, S. F.; Quan, Q.; Zhang, N.; Xu, Y. J.; Chen, H. M. Electrocatalysis for the Oxygen Evolution Reaction: Recent Development and Future Perspectives. *Chem. Soc. Rev.* **2017**, *46*, 337−365.

8. Tahir, M.; Pan, L.; Idrees, F.; Zhang, X.; Wang, L.; Zou, J.-J.; Wang, Z. L. Electrocatalytic Oxygen Evolution Reaction for Energy Conversion and Storage: A Comprehensive Review. *Nano Energy* **2017**, *37*, 136−157.

9. Cheng, W.; Lu, X. F.; Luan, D.; Lou, X. W. D. NiMn-Based Bimetal-Organic Framework Nanosheets Supported on Multi-Channel Carbon Fibers for Efficient Oxygen Electrocatalysis. *Angew. Chem., Int. Ed.* **2020**, *59*, 18234−18239.

10. Su, H.; Zhao, X.; Cheng, W.; Zhang, H.; Li, Y.; Zhou, W.; Liu, M.; Liu, Q. Hetero-N-Coordinated Co Single Sites with High Turnover Frequency for Efficient Electrocatalytic Oxygen Evolution in an Acidic Medium. *ACS Energy Lett.* **2019**, *4*, 1816−1822.

11. Zhang, B.; Zheng, X.; Voznyy, O.; Comin, R.; Bajdich, M.; Garcia-Melchor, M.; Han, L.; Xu, J.; Liu, M.; Zheng, L.; Garcia de Arquer, F. P.; Dinh, C. T.; Fan, F.; Yuan, M.; Yassitepe, E.; Chen, N.; Regier, T.; Liu, P.; Li, Y.; De Luna, P.; Janmohamed, A.; Xin, H. L.; Yang, H.; Vojvodic, A.; Sargent, E. H. Homogeneously Dispersed Multimetal Oxygen-evolving Catalysts. *Science* **2016**, *352*, 333−337.

12. Lim, B.; Jiang, M.; Camargo, P. H. C.; Cho, E. C.; Tao, J.; Lu, X.; Zhu, Y.; Xia, Y. Pd-Pt Bimetallic Nanodendrites with High Activity for Oxygen Reduction. *Science* **2009**, *324*, 1302−1305.

13. Greeley, J.; Stephens, I. E. L.; Bondarenko, A. S.; Johansson, T. P.; Hansen, H. A.; Jaramillo, T. F.; Rossmeisl, J.; Chorkendorff, I.; Nørskov, J. K. Alloys of Platinum and Early Transition Metals as Oxygen Reduction Electrocatalysts. *Nat. Chem.* **2009**, *1*, 552−556.

14. Zhang, T.; Li, S. C.; Zhu, W.; Zhang, Z. P.; Gu, J.; Zhang, Y. W. Shape-tunable Pt-Ir Alloy Nanocatalysts with High Performance in Oxygen Electrode Reactions. *Nanoscale* **2017**, *9*, 1154−1165.

15. Favaro, M.; Valero-Vidal, C.; Eichhorn, J.; Toma, F. M.; Ross, P. N.; Yano, J.; Liu, Z.; Crumlin, E. J. Elucidating the Alkaline Oxygen Evolution Reaction Mechanism on

Platinum. *J. Mater. Chem. A* **2017**, *5*, 11634−11643.

16. Kanan, M. W.; Nocera, D. G. In Situ Formation of an Oxygen-Evolving Catalyst in Neutral Water Containing Phosphate and $Co^{2+}$. *Science* **2008**, *321*, 1072−1075. [16]

17. McCrory, C. C. L.; Jung, S.; Peters, J. C.; Jaramillo, T. F. Benchmarking Heterogeneous Electrocatalysts for the Oxygen Evolution Reaction. *J. Am. Chem. Soc.* **2013**, *135*, 16977−16987.

18. Audichon, T.; Napporn, T. W.; Canaff, C.; Morais, C.; Comminges, C.; Kokoh, K. B. $IrO_2$ Coated on $RuO_2$ as Efficient and Stable Electroactive Nanocatalysts for Electrochemical Water Splitting. *J. Phys. Chem. C* **2016**, *120*, 2562−2573.

19. Naito, T.; Shinagawa, T.; Nishimoto, T.; Takanabe, K. Recent Advances in Understanding Oxygen Evolution Reaction Mechanisms over Iridium Oxide. *Inorg. Chem. Front.* **2021**, *8*, 2900−2917.

20. Cady, C.; Gardner, G.; Maron, Z.; Retuerto, M.; Go, Y. B.; Segan, S.; Greenblatt, M.; Dismukes, G. C. Tuning the Electro-catalytic Water Oxidation Properties of $AB_2O_4$ Spinel Nanocrystals: A (Li, Mg, Zn) and B (Mn, Co) Site Variants of $LiMn_2O_4$. *ACS Catal.* **2015**, *5*, 3403−3410.

21. Avcı, Ö. N.; Sementa, L.; Fortunelli, A. Mechanisms of the Oxygen Evolution Reaction on $NiFe_2O_4$ and $CoFe_2O_4$ Inverse-Spinel Oxides. *ACS Catal.* **2022**, *12*, 9058−9073.

22. Harada, M.; Kotegawa, F.; Kuwa, M. Structural Changes of Spinel $MCo_2O_4$ (M = Mn, Fe, Co, Ni, and Zn) Electrocatalysts during the Oxygen Evolution Reaction Investigated by In Situ X-ray Absorption Spectroscopy. *ACS Appl. Energy Mater.* **2022**, *5*, 278−294.

23. Bezerra, L. S.; Maia, G. Developing Efficient Catalysts for the OER and ORR using a Combination of Co, Ni, and Pt Oxides along with Graphene Nanoribbons and $NiCo_2O_4$. *J. Mater. Chem. A* **2020**, *8*, 17691−17705.

24. Spöri, C.; Falling, L. J.; Kroschel, M.; Brand, C.; Bonakdarpour, A.; Kühl, S.; Berger, D.; Gliech, M.; Jones, T. E.; Wilkinson, D. P.; Strasser, P. Molecular Analysis of the Unusual Stability of an $IrNbO_x$ Catalyst for the Electrochemical Water Oxidation to Molecular Oxygen (OER). *ACS Appl. Mater. Interfaces* **2021**, *13*, 3748−3761.

25. Friebel, D.; Louie, M. W.; Bajdich, M.; Sanwald, K. E.; Cai, Y.; Wise, A. M.; Cheng, M.-J.; Sokaras, D.; Weng, T.-C.; Alonso-Mori, R.; Davis, R. C.; Bargar, J. R.; Nørskov, J. K.; Nilsson, A.; Bell, A. T. Identification of Highly Active Fe Sites in (Ni,Fe)OOH for Electrocatalytic Water Splitting. *J. Am. Chem. Soc.* **2015**, *137*, 1305−1313.

26. Li, Y.-F.; Li, J.-L.; Liu, Z.-P. Structure and Catalysis of NiOOH: Recent Advances on Atomic Simulation. *J. Phys. Chem. C* **2021**, *125*, 27033−27045.

27. Ahn, H. S.; Bard, A. J. Surface Interrogation Scanning Electrochemical Microscopy of $Ni_{1-x}Fe_xOOH$ (0 < x < 0.27) Oxygen Evolving Catalyst: Kinetics of the "fast" Iron Sites. *J. Am. Chem. Soc.* **2016**, *138*, 313−318.

28. Stevens, M. B.; Trang, C. D. M.; Enman, L. J.; Deng, J.; Boettcher, S. W. Reactive Fe-Sites in Ni/Fe (Oxy)Hydroxide Are Responsible for Exceptional Oxygen Electrocatalysis Activity. *J. Am. Chem. Soc.* **2017**, *139*, 11361−11364.

29. Goldsmith, Z. K.; Harshan, A. K.; Gerken, J. B.; Vörös, M.; Galli, G.; Stahl, S. S.; Hammes-Schiffer, S. Characterization of NiFe Oxyhydroxide Electrocatalysts by Integrated Electronic Structure Calculations and Spectroelectrochemistry. *Proc. Natl. Acad. Sci. U.S.A.* **2017**, *114*, 3050−3055.

30. Diaz-Morales, O.; Ledezma-Yanez, I.; Koper, M. T. M.; Calle Vallejo, F. Guidelines for the Rational Design of Ni-Based Double Hydroxide Electrocatalysts for the Oxygen Evolution Reaction. *ACS Catal.* **2015**, *5*, 5380−5387.

31. Trześniewski, B. J.; Diaz-Morales, O.; Vermaas, D. A.; Longo, A.; Bras, W.; Koper, M. T.; Smith, W. A. In Situ Observation of Active Oxygen Species in Fe-Containing Ni-Based Oxygen Evolution Catalysts: The Effect of pH on Electrochemical Activity. *J. Am. Chem. Soc.* **2015**, *137*, 15112−15121.

32. Görlin, M.; Chernev, P.; Ferreira de Araújo, J.; Reier, T.; Dresp, S.; Paul, B.; Krähnert, R.; Dau, H.; Strasser, P. Oxygen Evolution Reaction Dynamics, Faradaic Charge Efficiency, and the Active Metal Redox States of Ni-Fe Oxide Water Splitting Electrocatalysts. *J. Am. Chem. Soc.* **2016**, *138*, 5603−5614.

33. Chen, J. Y. C.; Dang, L.; Liang, H.; Bi, W.; Gerken, J. B.; Jin, S.; Alp, E. E.; Stahl, S. S. Operando Analysis of NiFe and Fe Oxyhydroxide Electrocatalysts for Water Oxidation: Detection of $Fe^{4+}$ by Mössbauer Spectroscopy. *J. Am. Chem. Soc.* **2015**, *137*, 15090−15093.

34. Shin, H.; Xiao, H.; Goddard, W. A. In Silico Discovery of New Dopants for Fe-Doped Ni Oxyhydroxide ($N_{1-x}Fe_xOOH$) Catalysts for Oxygen Evolution Reaction. *J. Am. Chem. Soc.* **2018**, *140*, 6745−6748.

35. Xiao, H.; Shin, H.; Goddard, W. A. Synergy between Fe and Ni in the Optimal Performance of (Ni,Fe)OOH Catalysts for the Oxygen Evolution Reaction. *Proc. Natl. Acad. Sci. U. S. A.* **2018**, *115*, 5872−5877.

36. Martirez, J. M. P.; Carter, E. A. Unraveling Oxygen Evolution on Iron-Doped *β*-Nickel Oxyhydroxide: The Key Role of Highly Active Molecular-like Sites. *J. Am. Chem. Soc.* **2019**, *141*, 693−705.

37. Tian, B.; Shin, H.; Liu, S.; Fei, M.; Mu, Z.; Liu, C.; Pan, Y.; Sun, Y.; Goddard, W. A.; Ding, M. Double-Exchange-Induced in situ Conductivity in Nickel-Based

Oxyhydroxides: An Effective Descriptor for Electrocatalytic Oxygen Evolution. *Angew. Chem., Int. Ed.* **2021**, *60*, 16448−16456.

38. Conway, B. E.; Bockris, J. OM.; Ammar, I. A. The Dielectric Constant of the Solution in the Diffuse and Helmholtz Double Layers at a Charged Interface in Aqueous Solution. *Trans. Faraday Soc*. **1951**, *47*, 756−760.

39. Bockris, J. O.; V Devanathan, M. A.; Müller, K. On the Structure of Charged Interfaces. *Proc. R. Soc. London, Ser. A* **1963**, *274*, 55−79.

40. Huang, J.; Li, M.; Eslamibidgoli, M. J.; Eikerling, M.; Groß, A. Cation Overcrowding Effect on the Oxygen Evolution Reaction. *JACS Au* **2021**, *1*, 1752−1765.

41. Huang, B.; Myint, K. H.; Wang, Y.; Zhang, Y.; Rao, R. R.; Sun, J.; Muy, S.; Katayama, Y.; Corchado Garcia, J.; Fraggedakis, D.; Grossman, J. C.; Bazant, M. Z.; Xu, K.; Willard, A. P.; Shao-Horn, Y. Cation-Dependent Interfacial Structures and Kinetics for Outer-Sphere Electron-Transfer Reactions. *J. Phys. Chem. C* **2021**, *125*, 4397−4411.

42. Marcus, Y. Effect of Ions on the Structure of Water-Structure Making and Breaking. *Chem. Rev.* **2009**, *109*, 1346−1370.

43. Marcus, Y. Viscosity *B*-Coefficients, Structural Entropies and Heat Capacities, and the Effects of Ions on the Structure of Water. *Journal of Solution Chemistry* **1994**, *23*, 831−848.

44. Garcia, A. C.; Touzalin, T.; Nieuwland, C.; Perini, N.; Koper, M. T. M. Enhancement of Oxygen Evolution Activity of Nickel Oxyhydroxide by Electrolyte Alkali Cations. *Angew. Chem., Int. Ed.* **2019**, *58*, 12999−13003.

45. Rao, R. R.; Huang, B.; Katayama, Y.; Hwang, J.; Kawaguchi, T.; Lunger, J. R.; Peng, J.; Zhang, Y.; Morinaga, A.; Zhou, H.; You, H.; Shao-Horn, Y. pH- and cation-dependent water oxidation on rutile $RuO_2$ (110). *J. Phys. Chem. C* **2021**, *125*, 8195−8207.

46. van der Heijden, O.; Eggebeen, J. J. J.; Trzesniowski, H.; Deka, N.; Golnak, R.; Xiao, J.; van Rijn, M.; Mom, R. V.; Koper, M. T. M. $Li^+$ Cations Activate NiFeOOH for Oxygen Evolution in Sodium and Potassium Hydroxide. *Angew. Chem. Int. Ed.* **2024**, 63, e202318692.

47. Van der Ven, A.; Morgan, D.; Meng, Y. S.; Ceder, G. Phase Stability of Nickel Hydroxides and Oxyhydroxides. *J. Electrochem. Soc.* **2006**, *153*, A210.

48. Wehrens-Dijksma, M.; Notten, P. H. L. Electrochemical Quartz Microbalance Characterization of $Ni(OH)_2$-Based Thin Film Electrodes. *Electrochim. Acta* **2006**, *51*, 3609−3621.

49. Menezes, P. W.; Yao, S.; Beltran-Suito, R.; Hausmann, J. N.; Menezes, P. V.; Driess,

M. Facile Access to Active g - NiOOH Electrocatalyst for Durable Water Oxidation Derived From an Intermetallic Nickel Germanide Precursor. *Angew. Chem., Int. Ed.* **2020**, *60*, 4640−4647.

50. Liang, Q.; Brocks, G.; Bieberle-Hütter, A. Oxygen Evolution Reaction (OER) Mechanism under Alkaline and Acidic Conditions. *J. Phys. Energy*, **2021**, *3*, 026001.

51. Man, I. C.; Su, H.-Y.; Calle-Vallejo, F.; Hansen, H. A.; Martinez, J. I.; Inoglu, N. G.; Kitchin, J.; Jaramillo, T. F.; Nørskov, J. K.; Rossmeisl, J. Universality in Oxygen Evolution Electrocatalysis on Oxide Surfaces. *ChemCatChem* **2011**, *3*, 1159−1165.

52. Henkelman, G.; Arnaldsson, A.; Jónsson, H. A Fast and Robust Algorithm for Bader Decomposition of Charge Density. *Comput. Mater. Sci.* **2006**, *36*, 354−360.

53. Yu, M.; Trinkle, D. R. Accurate and Efficient Algorithm for Bader Charge Integration. *J. Chem. Phys.* **2011**, *134*, No. 064111.

54. Groß, A. Grand-Canonical Approaches to Understand Structures and Processes at Electrochemical Interfaces from an Atomistic Perspective. *Curr. Opin Electrochem* **2021**, *27*, 100684.

55. Chen, L. D.; Urushihara, M.; Chan, K.; Nørskov, J. K. Electric Field Effects in Electrochemical $CO_2$ Reduction. *ACS Catal.* **2016**, *6*, 7133−7139.

56. Tomasi, J.; Mennucci, B.; Cammi, R. Quantum Mechanical Continuum Solvation Models. *Chem. Rev.* **2005**, *105*, 2999−3093.

57. Huang, J.; Chen, S. L.; Eikerling, M. Grand-Canonical Model of Electrochemical Double Layers from a Hybrid Density−Potential Functional. *J. Chem. Theory Comput.* **2021**, *17*, 2417−2430.

58. Gongadze, E.; Iglič, A. Decrease of Permittivity of an Electrolyte Solution Near a Charged Surface due to Saturation and Excluded Volume Effects. *Bioelectrochemistry* **2012**, *87*, 199−203.

59. Nakayama, Y.; Andelman, D. Differential Capacitance of the Electric Double Layer: the Interplay Between Ion Finite Size and Dielectric Decrement. *J. Chem. Phys.* **2015**, *142*, 044706.

60. Chandra, A. Static Dielectric Constant of Aqueous Electrolyte Solutions: Is There Any Dynamic Contribution? *J. Chem. Phys.* **2000**, *113*, 903−905.

61. Lee, C.; McCammon, J. A.; Rossky, P. J. The Structure of Liquid Water at an Extended Hydrophobic Surface. *J. Chem. Phys.* **1984**, *80*, 4448−4455.

62. Toney, M. F.; Howard, J. N.; Richer, J.; Borges, G. L.; Gordon, J. G.; Melroy, O. R.; Wiesler, D. G.; Yee, D.; Sorensen, L. B. Voltage Dependent Ordering of Water Molecules at an Electrode-Electrolyte Interface. *Nature* **1994**, *368*, 444−446.

63. Shinagawa, T.; Garcia-Esparza, A. T.; Takanabe, K. Insight on Tafel Slopes from a

Microkinetic Analysis of Aqueous Electrocatalysis for Energy Conversion. *Sci. Rep.* **2015**, *5*, 13801.

64. Corrigan, D. A.; The Catalysis of the Oxygen Evolution Reaction by Iron Impurities in Thin Film Nickel Oxide Electrodes. *J. Electrochem. Soc.* **1987**, 134, 377

65. Rao, R. R.; Corby, S.; Bucci, A.; García-Tecedor, M.; Mesa, C. A.; Rossmeisl, J.; Giménez, S.; Lloret-Fillol, J.; Stephens, I. E. L.; Durrant, J. R. Spectroelectrochemical Analysis of the Water Oxidation Mechanism on Doped Nickel Oxides. *J. Am. Chem. Soc.* **2022**, 144, 7622-7633

66. Zhong, G.; Cheng, T.; Shah, A. H.; Wan, C.; Huang, Z.; Wang, S.; Leng, T.; Huang, Y.; Goddard, W. A.; Duan, X. Determining the Hydronium p$K_a$ at Platinum Surfaces and the Effect on pH Dependent Hydrogen Evolution Reaction Kinetics. *Proc. Natl. Acad. Sci. U. S. A.* **2022**, *119*, No. e2208187119.

67. Wang, X. S.; Xu, C. C.; Jaroniec, M.; Zheng, Y.; Qiao, S. Z. Anomalous Hydrogen Evolution Behavior in High-pH Environment Induced by Locally Generated Hydronium Ions. *Nat. Commun.* **2019**, 10, 4876.

68. Fukushima, T.; Fukasawa, M.; Murakoshi, K. Unveiling the Hidden Energy Profiles of the Oxygen Evolution Reaction via Machine Learning Analyses. *J. Phys. Chem. Lett.* **2023**, *14*, 6808−6813.

69. Suzuki, K.; Li, X.; Wang, Y.; Nagasawa, F.; Murakoshi, K. Active Intermediates in Plasmon-Induced Water Oxidation at Au Nanodimer Structures on a Single Crystal of $TiO_2$. *ACS Energy Lett.* **2020**, 5, 4, 1252–1259.

70. Suzuki, K.; Li, X.; Toda, T.; Nagasawa, F.; Murakoshi, K. Plasmon-Accelerated Water Oxidation at Ni-Modified Au Nanodimers on $TiO_2$ Single Crystals. *ACS Energy Lett.* **2021**, 6, 12, 4374–4382.

71. Kresse, G.; Furthmüller, J. Efficient Iterative Schemes for Ab Initio Total-Energy Calculations Using a Plane-Wave Basis Set. *Phys. Rev. B* **1996**, *54*, 11169−11186.

72. Kresse, G.; Furthmüller, J. Efficiency of Ab-Initio Total Energy Calculations for Metals and Semiconductors Using a Plane-Wave Basis Set. *Comput. Mater. Sci.* **1996**, *6*, 15−50.

73. Blöchl, P. E. Projector Augmented-Wave Method. *Phys. Rev. B* **1994**, *50*, 17953.

74. Perdew, J. P.; Burke, K.; Ernzerhof, M. Generalized Gradient Approximation Made Simple. *Phys. Rev. Lett.* **1996**, *77*, 3865−3868.

75. Dudarev, S. L.; Botton, G. A.; Savrasov, S. Y.; Humphreys, C. J.; Sutton, A. P. Electron-Energy-Loss Spectra and the Structural Stability of Nickel Oxide: An LSDA+U Study. *Phys. Rev. B* **1998**, *57*, 1505−1509.

76. Li, Y. F.; Selloni, A. Mechanism and Activity of Water Oxidation on Selected Surfaces

of Pure and Fe-Doped NiO$_x$. *ACS Catal.* **2014**, *4*, 1148−1153.

77. Tkalych, A. J.; Yu, K.; Carter, E. A. Structural and Electronic Features of *β*-Ni(OH)$_2$ and *β*-NiOOH from First Principles. *J. Phys. Chem. C* **2015**, *119*, 24315−24322.

78. Li, Y.-F.; Selloni, A. Mosaic Texture and Double c-Axis Periodicity of beta-NiOOH: Insights from First-Principles and Genetic Algorithm Calculations. *J. Phys. Chem. Lett.* **2014**, *5*, 3981−3985.

79. Grimme, S.; Antony, J.; Ehrlich, S.; Krieg, H. A Consistent and Accurate *Ab Initio* Parametrization of Density Functional Dispersion Correction (DFT-D) for the 94 Elements H-Pu. *J. Chem. Phys.* **2010**, *132*, 154104.

80. McQuarrie, D. A.; Simon, J. D. *Molecular Thermodynamics*; University Science Books: Herndon, VA, 1999.

81. Lyalin, A.; Nakayama, A.; Uosaki, K.; Taketsugu, T. Functionalization of Monolayer h-BN by a Metal Support for the Oxygen Reduction Reaction. *J. Phys. Chem. C* **2013**, *117*, 21359− 21370.

82. Lyalin, A.; Nakayama, A.; Uosaki, K.; Taketsugu, T. Theoretical Predictions for Hexagonal BN Based Nanomaterials as Electrocatalysts for the Oxygen Reduction Reaction. *Phys. Chem. Chem. Phys.* **2013**, *15*, 2809−2820.

83. Mathew, K.; Sundararaman, R.; Letchworth-Weaver, K.; Arias, T.; Hennig, R. G. Implicit Solvation Model for Density-Functional Study of Nanocrystal Surfaces and Reaction Pathways. *J. Chem. Phys.* **2014**, *140*, 084106.

84. VASPsol: Implicit Solvation and Electrolyte Model for Density-Functional Theory. https://github.com/henniggroup/VASPsol, 2021.

85. Bajdich, M.; García-Mota, M.; Vojvodic, A.; Nørskov, J. K.; Bell, A. T. Theoretical Investigation of The Activity of Cobalt Oxides for The Electrochemical Oxidation of Water. *J. Am. Chem. Soc.* **2013**, *135*, 13521−13530.

86. Lu, Z.; Chen, G.; Li, Y.; Wang, H.; Xie, J.; Liao, L.; Liu, C.; Liu, Y.; Wu, T.; Li, Y.; et al. Identifying the Active Surfaces of Electrochemically Tuned LiCoO$_2$ for Oxygen Evolution Reaction. *J. Am. Chem. Soc.* **2017**, *139*, 6270−6276.

87. Monkhorst, H. J.; Pack, J. D. Special Points for Brillouin-Zone Integrations. *Phys. Rev. B* **1976**, *13*, 5188−5192.

88. Momma, K.; Izumi, F. Vesta 3 for Three-Dimensional Visualization of Crystal, Volumetric and Morphology Data. *J. Appl. Crystallogr.* **2011**, *44*, 1272−1276.

# Supplementary Materials

# Directing the Electrode-Electrolyte Interface Towards Active Nickel-Based Electrocatalysts for Oxygen Evolution Reaction

Ben Wang[1], Tomohiro Fukushima[1], Hiro Minamimoto[2], Andrey Lyalin[1,3,†], Kei Murakoshi[1,‡], Tetsuya Taketsugu[1,4,#]

[1]Department of Chemistry, Faculty of Science, Hokkaido University, Sapporo 060-0810, Japan

[2]Department of Chemical Science and Engineering, Graduate School of Engineering, Kobe University, Kobe 657-8501, Japan

[3]Research Center for Energy and Environmental Materials (GREEN), National Institute for Materials Science, Namiki 1-1, Tsukuba 305-0044, Japan

[4]Institute for Chemical Reaction Design and Discovery (WPI-ICReDD), Hokkaido University, Sapporo 001-0021, Japan

[†]Email: lyalin@icredd.hokudai.ac.jp
[‡]Email: kei@sci.hokudai.ac.jp
[#]Email: t-taketsugu@eis.hokudai.ac.jp

**OER energetics**

To describe the energetics of the OER process we use the concept of the computational hydrogen electrode (CHE) described by Nørskov et al.,[1,2] where the electrochemical potential of the proton-electron pair at the given electrode potential is expressed as following[3,4]:

$$\mu(H^+) + \mu(e^-) = \frac{1}{2}\mu^0(H_2) - eU_{SHE} - k_B T \ln(10)\text{pH} = \frac{1}{2}\mu^0(H_2) - eU_{RHE}. \quad (S1)$$

Here $\mu(e^-)$ is the electrochemical potential of the electron and $\mu^0(H_2)$ is the chemical potential of hydrogen gas under standard conditions ($T$ = 298.15 K, $p_{H_2}$ = 1 bar), while $U_{SHE}$ and $U_{RHE}$ are the potentials on the standard and reversible hydrogen electrode scales, respectively. Assuming equilibrium $H_2O(l) \leftrightarrow H^+ + OH^-$, one can obtain the chemical potential of the $OH^-$ - electron pair:

$$\mu(OH^-) - \mu(e^-) = \mu^0(H_2O) - \frac{1}{2}\mu^0(H_2) + eU_{RHE}, \quad (S2)$$

where $\mu^0(H_2O)$ is the chemical potential of liquid water under standard conditions ($T$ = 298.15 K, $p_{H_2O}$ = 0.035 bar) and $\mu(OH^-)$ is the electrochemical potential of hydroxyl anion $OH^-$.[3,4]

Equation (S2) allows us to calculate the Gibbs free energy changes $\Delta G_n$ for each reaction step $n$ of the conventional OER pathway under alkaline conditions (2a) - (2d)[4]:

$$\Delta G_{1,\text{single}} = E(*OH) - E(*) - E(H_2O) + \frac{1}{2}E(H_2) + \Delta E_{ZPE,1} - T\Delta S_1 - eU_{RHE}, \quad (S3a)$$

$$\Delta G_{2,\text{single}} = E(*O) - E(*OH) + \frac{1}{2}E(H_2) + \Delta E_{ZPE,2} - T\Delta S_2 - eU_{RHE}, \quad (S3b)$$

$$\Delta G_{3,\text{single}} = E(*OOH) - E(*O) - E(H_2O) + \frac{1}{2}E(H_2) + \Delta E_{ZPE,3} - T\Delta S_3 - eU_{RHE}, \quad (S3c)$$

$$\Delta G_{4,\text{single}} = E(*) - E(*OOH) + 2E(H_2O) - \frac{3}{2}E(H_2) + \Delta G_0 + \Delta E_{ZPE,4} - T\Delta S_4 - eU_{RHE}, \quad (S3d)$$

where $E(*)$, $E(*M)$, $E(H_2O)$, and $E(H_2)$ represent the total energy of the pure surface slab, the surface with the adsorbed species M = (OH, O, OOH), $H_2O$, and $H_2$ molecules as it follows from DFT calculations, respectively. $\Delta E_{ZPE,n}$, and $\Delta S_n$ are changes in zero-point energies (ZPE) and entropies for each reaction step $n$, $T$ is a temperature, and $\Delta G_0$ = 4.92 eV is the overall free energy difference of the reaction. Note that expression (5d) was obtained using assumption $\mu^0(O_2) = 4.92$ eV + 2 ($\mu^0(H_2O) - \mu^0(H_2)$) to avoid the well-known inaccuracy in the DFT value of the energy of $O_2$ molecule.[1,5,6]

The changes of free energies $\Delta G_n$ in Eqs. (S3a) - (S3d) satisfy the following expression[4]:

$$\sum_n \Delta G_n = \Delta G_0 - 4eU_{RHE}, \tag{S4}$$

while the overpotential is given by

$$\eta = \frac{1}{e} \max_n [\Delta G_n] - U_0, \tag{S5}$$

where $U_0 = \Delta G_0/4e = 1.23$ V is the equilibrium potential. Thus, for the ideal catalyst, all four reaction steps are equal: $\Delta G_{1,single} = \Delta G_{2,single} = \Delta G_{3,single} = \Delta G_{4,single} = 1.23$ eV, which corresponds to zero overpotential $\eta = 0$. However, for the real catalyst, the adsorption energies of OER intermediates are unbalanced, and the overpotential $\eta$ is required to drive the step with the largest $\Delta G_{n,single}$.

The corresponding changes in free energy $\Delta G_n$ ($n=1,\ldots,5$) along the dual-site reaction pathway (3a) – (3e) can be calculated as follows:

$$\Delta G_{1,dual} = E(*OH + \star H_2O) - E(* + \star OH \cdots H^O) - E(H_2O) + \frac{1}{2}E(H_2) +$$
$$+ \Delta E_{ZPE,1} - T\Delta S_1 - eU_{RHE}, \tag{S6a}$$

$$\Delta G_{2,dual} = E(*OH + \star OH) - E(*OH + \star H_2O) + \frac{1}{2}E(H_2) +$$
$$+ \Delta E_{ZPE,2} - T\Delta S_2 - eU_{RHE}, \tag{S6b}$$

$$\Delta G_{3,dual} = E(*O \cdots H_2O^{(*,O)} + \star OH) - E(*OH + \star OH) - E(H_2O) + \frac{1}{2}E(H_2) +$$
$$+ \Delta E_{ZPE,3} - T\Delta S_3 - eU_{RHE}, \tag{S6c}$$

$$\Delta G_{4,dual} = E(*OOH + \star OH \cdots H^O) - E(*O \cdots H_2O^{(*,O)} + \star OH) + \Delta E_{ZPE,4} - T\Delta S_4 \tag{S6d}$$

$$\Delta G_{5,dual} = E(* + \star OH \cdots H^O) - E(*OOH + \star OH \cdots H^O) + 2E(H_2O) - \frac{3}{2}E(H_2) + \Delta G_0 +$$
$$+ \Delta E_{ZPE,5} - T\Delta S_5 - eU_{RHE}. \tag{S6e}$$

Expressions (S6a)−(S6e) satisfy conditions (S4) and (S5) with $n=1,\ldots,5$. Note that $\Delta G_4$ in step (S6d) is negative and does not depend on the potential, as it is the chemical transition to the energetically favorable configuration.

**Kinetics modeling**

The electrical current through an electrode for the simple unimolecular one-electron redox reaction, considering that both a cathodic and an anodic reaction occur on the same electrode can be described by the Butler-Volmer equation:

$$J = J_0 \{e^{\alpha_a f \eta} - e^{-\alpha_c f \eta}\} \tag{S7}$$

where $J$ is the current density in A/m², $J_0$ is the exchange current density, $\eta$ is the overpotential defined as a difference between the electrode and the standard potential ($\eta = E - E_0$), $f$ denotes the ratio $F/RT$, where $F$ is the Faraday's constant ($F = 96485$ C mol$^{-1}$), $R$ is the universal gas constant ($R = 8.314$ J mol$^{-1}$ K$^{-1}$), and $T$ is the temperature, while $\alpha_a$ and $\alpha_c$ are the so-called anodic and cathodic dimensionless charge transfer coefficients, generally related by $\alpha_a + \alpha_c = 1$. For the sake of simplicity, we adopt the commonly used assumption $\alpha_a = \alpha_c = \alpha = 0.5$.

Let us consider the kinetics of the oxygen evolution reaction. In this case, the kinetic rate constant for the forward (+) and backward (-) $i$th reaction step depends on the applied potential as follows:

$$k_i = k_i^0 e^{(1-\alpha) f \eta_i} \text{ and } k_{-i} = k_{-i}^0 e^{-\alpha f \eta_i} \tag{S8}$$

where $k_i^0$ defined the standard rate constant for the corresponding $k_i$, and $\eta_i$ is the overpotential for the $i$th reaction step.

In the case of the single-site reaction mechanism (**Figure 1c**) we follow the formalism described in Ref. [7].

When the first step (2a) determines the overall reaction, the corresponding reaction rate can be expressed as follows:

$$r_{2a} = k_{2a}\, \theta_0\, a_{OH^-} = k_{2a} a_{OH^-} \tag{S9}$$

where $\theta_0$ is the surface coverage by the empty site ($\theta_0 = 1$, when all sites are empty), $k_{2a}$ is the rate constant for equation (2a), and $a_{OH^-}$ is the hydroxide ion activity (effective concentration of the hydroxide ion in the solution).

The kinetic current related to the $i$th reaction step is defined by the reaction rate:

$$I_i = n\, F\, S\, r_i, \tag{S10}$$

where $S$ denotes the surface area of the catalyst and $n$ is the number of electrons involved.

Therefore, the kinetic current density when (2a) determines the overall reaction rate is expressed as follows:

$$J = nFA\, r_{2a} = nFA\, k_{2a}^0 a_{OH^-}\, e^{(1-\alpha)f\eta_{2a}} \tag{S11}$$

where $A$ is the reaction site density of the catalytic surface $S$.

When the second step (2b) determines the overall reaction, reaction (2a) should be at equilibrium:

$$r_{2a} = r_{-2a}\,. \tag{S12}$$

The reaction rate for the backward reaction can be expressed as follows:

$$r_{-2a} = k_{-2a}\, \theta_1 \tag{S13}$$

where $\theta_1$ is the surface coverage by the *OH site.

Taking into account (S8), (S9), (S12) and (S13) one can get

$$\theta_1 = \widehat{K}_{2a}^0\, e^{f\eta_{2a}}\, a_{OH^-}\, \theta_0 \tag{S14}$$

where $\widehat{K}_i^0$ is the ratio of the standard forward and backward rate constants for the $i$th reaction: $\widehat{K}_i^0 = k_i^0/k_{-i}^0$.

As reaction (2b) is the limiting step the catalytic surface can be covered either by the empty or by the *OH sites, resulting in the following condition:

$$\theta_0 + \theta_1 = 1\,. \tag{S15}$$

Therefore, one can get the following expression for the potential dependent coverage $\theta_1$:

$$\theta_1 = \frac{\widehat{K}_{2a}^0\, e^{f\eta_{2a}}\, a_{OH^-}}{1+\widehat{K}_{2a}^0\, e^{f\eta_{2a}}\, a_{OH^-}}\,. \tag{S16}$$

The expression (S16) allows us to calculate the kinetic current density when (2b) determines the overall reaction rate is expressed as follows:

$$J = nFA\, r_{2b} = nFA\, \theta_1 k_{2b}^0 a_{OH^-}\, e^{(1-\alpha)f\eta_{2b}}\,. \tag{S17}$$

Let us now consider the OER kinetics for the dual-site reaction mechanism. Our calculations demonstrate that in this case, the (3c) reaction is the rate-limiting step. To describe the overall kinetics in the case of the dual-site mechanism we can use a similar approach as for the single-site, taking into account that there are two different adsorption sites A and B. According to the scheme **Figure 1d**, electrochemical reactions (3a), (3c), and (3e) occur on the site A, while the reaction (3b) occurs on the site B.

The first step in the dual-site mechanism process (3a) is $OH^-$ adsorption on the empty site A, with the initial coverage by the empty site A, $\theta_0^A = 1$, similar to the (2a) process, accompanied by the reorganization of the site B. The kinetic current in this case can be described by an expression similar to equation (S11), where additional factor $\gamma'$ accounts for reorganization of the site B.

$$J = nFA\, \gamma' k_{3a}^0 a_{OH^-}\, e^{(1-\alpha)f\eta_{3a}} \tag{S18}$$

The second step in the dual-site mechanism process (3b) is $OH^-$ adsorption on the site B, accompanied by $H_2O$ desorption from the site B. This can be described by the following equation, where factor $\gamma''$ accounts for $H_2O$ desorption from site B:

$$J = nFA\, \gamma'' k_{3b}^0 a_{OH^-}\, e^{(1-\alpha)f\eta_{3b}} \tag{S19}$$

In the third step (3c) of the dual-site reaction mechanism, the third $OH^-$ anion attacks the *OH site A leading to its deprotonation and formation of the water molecule. Therefore, this step is similar to the (2b) process in the single-site mechanism, and can be described by the following expression:

$$J = nFA\, \theta_1^A\, k_{3c}^0 a_{OH^-}\, e^{(1-\alpha)f\eta_{3c}} \tag{S20}$$

where coverage $\theta_1^A$ of the site A given by the following expression:

$$\theta_1^A = \frac{\hat{K}_{3a}^0\, e^{f\eta_{3a}}\, a_{OH^-}}{1+\hat{K}_{3a}^0\, e^{f\eta_{3a}}\, a_{OH^-}} \tag{S21}$$

Here, we assume for simplicity that $\gamma' \approx \gamma'' \approx 1$ and reactions on the sites A and B are independent.

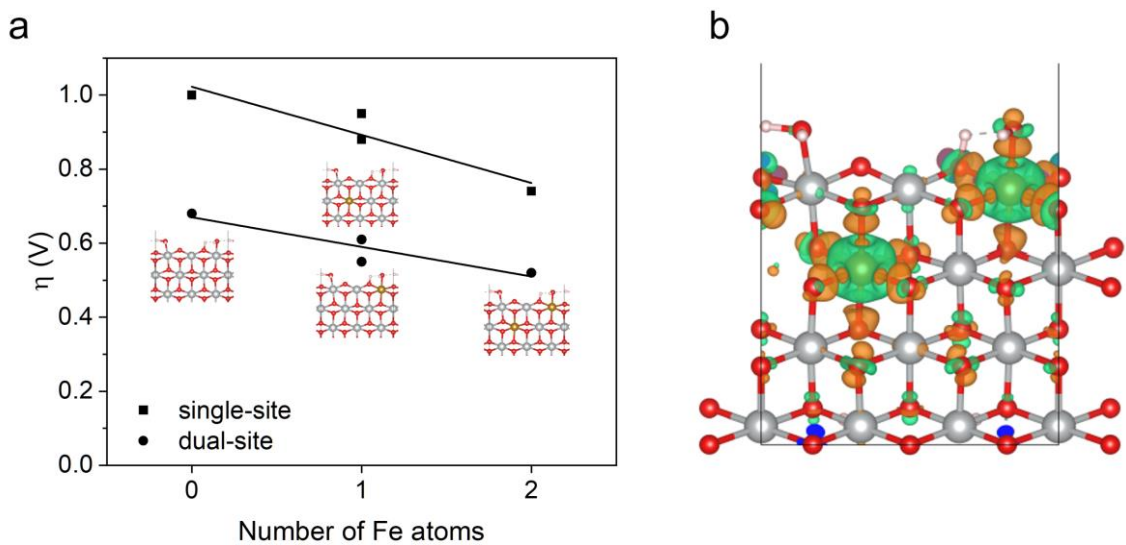

**Figure S1.** (a) Overpotential as a function of Fe concentration for the single-site (squares) and the dual-site (dots) reaction mechanism. Positions of Fe atoms in a unit cell are shown in inserts. (b) Charge density difference upon incorporation of two Fe impurities. Green and yellow represent the depletion and accumulation of electrons, respectively. The charge density is plotted with an isosurface value of 0.04 e Å$^{-3}$. Ni atoms are colored grey, Fe atoms are gold, O atoms are red, and H atoms are light pink

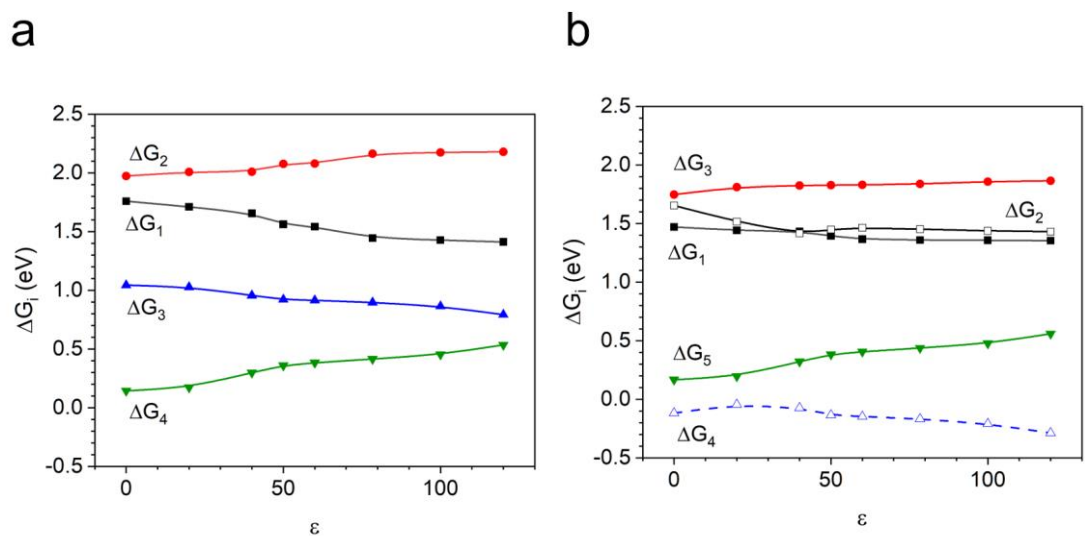

**Figure S2. Changes in free energy as a function of dielectric constant**. Changes in free energy Δ$G_n$ for each elementary step along the reaction pathway at 0 $V_{RHE}$, calculated for the single-site (left) and dual-site (right) mechanisms, respectively. Two Fe atom doped γ-NiOOH (100) surface is considered.

# References

[1] Nørskov, J. K.; Rossmeisl, J.; Logadottir, A.; Lindqvist, L.; Kitchin, J. R.; Bligaard, T.; Jónsson, H. Origin of the Overpotential for Oxygen Reduction at a Fuel-Cell Cathode. *J. Phys. Chem. B* **2004**, *108*, 17886−17892.

[2] Peterson, A. A.; Abild-Pedersen, F.; Studt, F.; Rossmeisl, J.; Nørskov, J. K. How Copper Catalyzes the Electroreduction of Carbon Dioxide into Hydrocarbon Fuels. *Energy Environ. Sci.* **2010**, *3*, 1311−1315.

[3] Huang, J.; Li, M.; Eslamibidgoli, M. J.; Eikerling, M.; Groß, A. Cation Overcrowding Effect on the Oxygen Evolution Reaction. *JACS Au* **2021**, *1*, 1752−1765.

[4] Liang, Q.; Brocks, G.; Bieberle-Hütter, A. Oxygen Evolution Reaction (OER) Mechanism under Alkaline and Acidic Conditions. *J. Phys. Energy*, **2021**, *3*, 026001.

[5] Man, I. C.; Su, H.-Y.; Calle-Vallejo, F.; Hansen, H. A.; Martinez, J. I.; Inoglu, N. G.; Kitchin, J.; Jaramillo, T. F.; Nørskov, J. K.; Rossmeisl, J. Universality in Oxygen Evolution Electrocatalysis on Oxide Surfaces. *ChemCatChem* **2011**, *3*, 1159−1165.

[6] Rossmeisl, J.; Logadottir, A.; Nørskov, J. K. Electrolysis of Water on (Oxidized) Metal Surfaces. *Chem. Phys.* **2005**, *319*, 178−184.

[7] Shinagawa, T.; Garcia-Esparza, A. T.; Takanabe, K. Insight on Tafel Slopes from a Microkinetic Analysis of Aqueous Electrocatalysis for Energy Conversion. *Sci. Rep.* **2015**, *5*, 13801